\documentclass{elsart}
\usepackage{natbib}
\usepackage{amssymb}
\usepackage{graphicx}
\newcommand{\be}  { \begin{equation} }
\newcommand{\ee}  { \end{equation}   }
\newcommand{\bea} { \begin{eqnarray} }
\newcommand{\eea} { \end{eqnarray}   }

\newcommand{\msun}{ \ensuremath{M_{\odot}  } }
\newcommand{\rmag}{ \ensuremath{r_{\rm m}  } }
\newcommand{\rco} { \ensuremath{r_{\rm co} } }

\newcommand{\mdot}{ \ensuremath{{\dot M}   } }

\newcommand{\bbag}{ \ensuremath{{\bar B}   } }

\newcommand{\pdot}{ \ensuremath{\dot P     } }

\newcommand{\Omegadot}{ \ensuremath{\dot \Omega   } }

\newcommand{\Bc}  { \ensuremath{B_{\rm c}  } }

\newcommand{\me}  { \ensuremath{m_{\rm e}  } }
\newcommand{\Omegak}{ \ensuremath{\Omega_\mathrm{K}} }
\newcommand{\mdotin}{ \ensuremath{{\dot M}_\mathrm{in}   } }
\newcommand{\mdotout}{ \ensuremath{{\dot M}_\mathrm{out}   } }
\newcommand{\rhon}{ \ensuremath{\rho_{\rm n}} }

\begin{document}
\begin{frontmatter}
\title{``High-field'' pulsars torqued by accretion disk?}
\author{Y. L. Yue},
\author{W. W. Zhu},
\author{R. X. Xu\corauthref{cor1}}
\corauth[cor1]{Corresponding author.}
\ead{r.x.xu@pku.edu.cn}
\ead[url]{http://vega.bac.pku.edu.cn/rxxu/}
\address{Astronomy Department, School of Physics, Peking University,
         Beijing 100871, China}

\begin{abstract}
The nature of AXPs/SGRs (anomalous X-ray pulsars/soft $\gamma$-ray
repeaters) and high field radio pulsars is still unclear even in
the magnetar and/or accretion models.
The detection of radio emission from AXP XTE J1810$-$197 and the
discovery of a debris disk around AXP 4U 0142+61 might shed light
on the problem.
We propose that AXPs/SGRs could be pulsars that have magnetic field
$ B\lesssim 10^{13} \mathrm{~G}$ as normal pulsars,
but in accretion environments.
We investigate these issues under the accretion model and find
that two of the AXPs/SGRs might be low mass quark stars if all
AXPs and SGRs are likely grouped together.
\end{abstract}

\begin{keyword}
 dense matter \sep
 pulsars: general \sep
% pulsars: individual (PSR ) \sep
 stars: neutron
\PACS
97.60.Gb \sep %Pulsars
97.60.Jd %Neutron stars
\end{keyword}

\end{frontmatter}

\section{Introduction}

As the number of observed pulsar increasing, there turns to be
some pulsars have apparent magnetic field strength
$\gtrsim10^{14}\mathrm{~G}$. These ``high field'' pulsars
(AXPs/SGRs and high field radio pulsars) challenge the physics of
matter in strong magnetic field, even being greater than the
critical magnetic field $\Bc = \me^2c^3/(e\hbar) \simeq
4.4\times10^{13} \mathrm{~G}$. The high field radio pulsars can
have magnetic fields as high as that of AXPs (two AXPs even have
$B$-field being stronger than the lowest field of AXPs).
Meanwhile, there are several kinds of pulsars that spin at periods
($P$) being similar to that of AXPs/SGRs ($P\sim 10\mathrm{~s}$),
such as DTNs (dim thermal neutron stars), part of RRATS
\citep[rotating radio transient sources,][]{McLaughlin06} and the
peculiar pulsar PSR J2144$-$3933 which seems under the death line
\citep{Young99}. A question rises then: why these pulsar-like
stars manifest so differently? The detection of radio emission
from AXP XTE J1810$-$197 \citep{Camilo06} and the discover of a
debris disk around AXP 4U 0142+61 \citep{Wang06} may show the
missing links between AXPs/SGRs and radio pulsars together.

AXPs/SGRs should have very strong magnetic field if their spindown
is torqued dominantly by magnetospheric activity. However, whether
the fields are really so strong or not is still under debating.
There are generally three kinds of models. (i) the magnetar model:
the sources to be actually neutron stars with ultra strong
magnetic fields $B$ of $10^{14}$--$10^{15}$ G
\citep[e.g.][]{Duncan92}; (ii) the accretion model: to be normal
pulsars with $B \lesssim 10^{13}$ G but with accretion disks
\citep[e.g.][]{Alpar01}; (iii) the hybrid model: to be neutron
stars with ultra strong magnetic fields but also surrounded by
accretion disks \citep[e.g.][]{Eksi03}.

The detection of radio emission \citep{Camilo06} from AXP and IR
emission from debris disk \citep{Wang06} around AXP  may favor the
accretion disk model. In this article, we investigate the nature
of AXPs/SGRs in the regime of accretion model of normal magnetic
fields $B\lesssim 10^{13} \mathrm{~G}$, considering the sources to
be possibly both neutron stars and/or quark stars. We find that
the two of the AXPs/SGRs might be low mass quark stars under
certain assumptions.

\section{AXPs/SGRs as accreting pulsars}

In the accretion model, the accretion rate $\mdot$ consists two
parts: $\mdotin$ which produces the X-ray emission, and $\mdotout$
which results in the accretion torque, i.e. $\mdot=\mdotin +
\mdotout$. Since the X-ray luminosity from magnetosphere of
rotation-powered X-ray pulsars is relatively low, we could assume
that the X-ray radiation would be accretion-originated,
\be
 L_\mathrm{x} = \eta \mdot_\mathrm{in} c^2 ,
\ee
where $\eta$ is the matter-to-energy conversion efficiency
and $c$ is the speed of light. If
the X-ray luminosity comes from release of gravitational energy in
case of a normal neutron star ($M\sim1.4\msun$), $\eta \sim
2GM/c^2/R \sim 10\%$, where $G$ is the gravitational constant
$M$ is the stellar mass and $R$ is the stellar radius;
 while if it comes from the phase transition
energy from baryons to quarks in case of quark stars (even with
low masses, \citep{xu05mn}), $\eta \sim \mathrm{100 ~MeV/930 ~MeV}
\sim10\%$. For massive quark stars ($M\sim1.4\msun$), both
gravitational energy and phase transition energy should be
considered, i.e., $\eta\sim20\%$. The difference is only by a
factor of 2. Considering the uncertainty of the estimated
efficiency, we use $\eta=10\%$ for all the cases below.

For a quark star with mass smaller than $1.4 \msun$, the magnetic
field could be expressed as,
\be
B \simeq 6.4\times10^{19}(P\pdot)^{1/2}
(\frac{M}{1.4\msun})^{1/2} (\frac{R}{10 ~\mathrm{km}})^{-2}~\mathrm{G},
\label{eq:B}
\ee
where $P$ is the spin period and $\pdot$ is the period derivative.
The density of low mass quark stars could almost be a constant
\citep{Alcock86},
 $\rho \sim 4\bbag$,
where $\bbag\sim 60$--$110 \mathrm{~MeV~fm^{-3}}$ ($1.1$--$2.0
\times10^{14} \mathrm{~g~cm^{-3}}$) is the bag constant in the MIT
bag model. The median value $\rho\sim 6 \times 10^{14}$ is about 2
times the nuclear saturation density $\rhon$ ($=2.7\times10^{14}
\mathrm{~g~cm^{-3}}$), and is also close to the mean density of a
normal neutron star with $M = 1.4 \msun$ and $R = 10 \mathrm{~km}$
($\bar{\rho} = 6.7\times10^{14}  \mathrm{~g~cm^{-3}} $). Thus we
may approximate the density $\sim 4{\bar B}$ for a $1.4\msun$
neutron star. The magnetic field could then be
\be B \simeq 6.4\times10^{19}(P\pdot)^{1/2} (\frac{R}{10
~\mathrm{km}})^{-1/2} \mathrm{~G}. \label{eq:B2} \ee The radius
$R$ could be much smaller than 10 km if pulsar-like stars are
actually quark stars (see, e.g., \citet{xu06asr,xu06cjaa} for more
backgrounds and discussion about quark stars).

Though the accretion torque in a propeller phase is still not
certain yet, it could also be summarized as
\citep[e.g.][]{jiangli05} \be T = -\mdotout \rmag^2
\Omegak(\rmag)[\frac{\Omega}{\Omegak(\rmag)}]^\gamma,
\label{eq:gamma} \ee where $\rmag$ is the magnetospheric radius
(Alfv\'en radius), \be
 r_{\rm m}\simeq({{B^2 R^6} \over { {\dot M}\sqrt{2GM} }})^{2/7},
\ee
$\Omega=2\pi/P$, $\Omegak(\rmag)$ is the Keplerian angular velocity at $\rmag$
and $\gamma$= -1 to 2 corresponding to different accretion models.
Another way of parameterizations proposed by \citet{Menou99} is
\be
T = -2\mdot \rmag^2 \Omegak(\rmag)[1-\frac{\Omega}{\Omegak(\rmag)}].
\label{eq:Menou}
\ee

\begin{figure}
\begin{center}
\includegraphics[width=.7\textwidth]{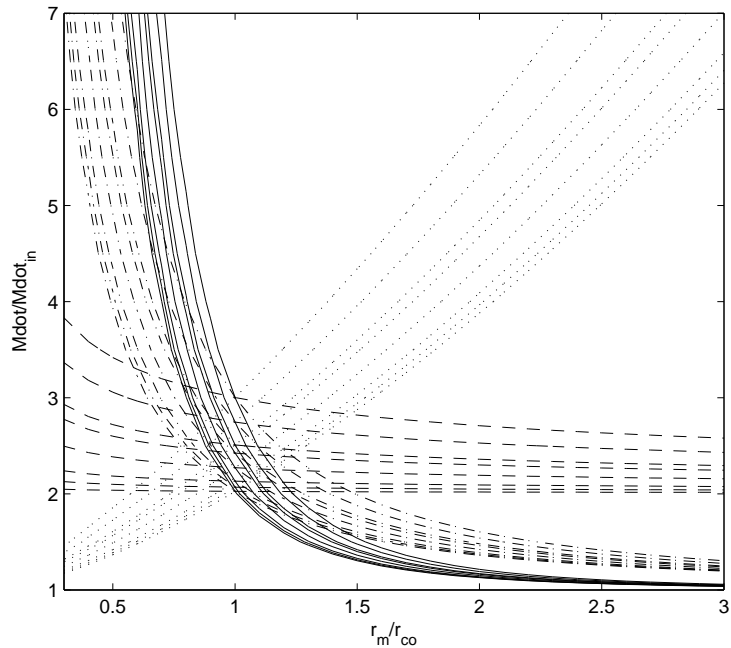}
\end{center}
\caption{ The parameter space for $\gamma=$ $-1$ (dotted), 0
(dashed), 1 (dash-dotted), and 2 (solid). We draw the ratio of
total accretion rate ($\dot M$) to the rate of accreting onto the
star's surface (${\dot M}_{\rm in}$) as a function of the ratio of
the Alfv\'en radius ($r_{\rm m}$) to the co-rotating radius
($r_{\rm co}$). Every single line represents a pulsar. On the
line, theoretical value of $\pdot$ equals to observational one.
For $\rmag/\rco\sim 1$, all the accretion torque cluster together.
In other region, the four models are quite different. In this
figure, SGR 1099+14 and AXP 1E 1048.1-5937 are in quark star
model, with $R=5$ km and $R=$ 2 km respectively in order to make
all the stars appear in a likely same group.
For all the available points on the lines, we have $\lesssim10^{13}$ G.
\label{fig:1}}
\end{figure}

Here we only use Eq. (\ref{eq:gamma}) since Eq. (\ref{eq:Menou})
does not give information about $\mdotin$ and
cannot be connected with $L_{\rm x}$.
In our calculation, we consider both angular momentum added from $\mdotin$
and angular momentum lost from $\mdotout$, thus
\be
T = I\Omegadot
= \mdotin \rmag^2 \Omegak(\rmag)
-\mdotout \rmag^2 \Omegak(\rmag)[\frac{\Omega}{\Omegak(\rmag)}]^\gamma.
\label{eq:odot}
\ee
In this case, $\rmag$ should around co-rotating radius
$\rco$ ($=[GMP^2/(4\pi^2)]^{1/3}$) and $\mdotout$ should
be of the same order of $\mdotin$. We use $\rmag/\rco$ as a free
parameter in the range 0.3--3 and $\mdot/\mdotin$ as another free
parameter in the range 1--7. On the grid, we calculate the
theoretical value of $\Omegadot$ from Eq. (\ref{eq:odot}) and
compare it with the observational values. The data are from
\citet{Woods06}. We use the larger value if the value of a
parameter (e.g. $\pdot$) varies in a range. There is 8 available
stars that has $P$, $\pdot$, $L_{\rm x}$, black-body temperature
and distance simultaneously. The results are presented in Fig.
(\ref{fig:1}). We find that two of the AXs/SGRs could be low-mass
quark stars in order to have all the stars in a likely same group.
If we apply $1.4\msun$ and 10 km for all the stars, these two
would not be grouped into that of others. The radius of SGR
1099+14 could be $R\sim$ 5 km and that of AXP 1E 1048.1-5937 could
be $R\sim$ 2 km, respectively. These values should not be very
exact since the values could vary in a certain range in order to
group all the stars. The uncertainty is about 1 km.
We have $B\lesssim10^{13}$ G at all available points in the parameter space.

\section{Conclusion and discussion}

We discussed the possible accretion environments around the long
period pulsars and the supports from observations. They could be
understood in a uniform accretion model, in which the requirement
of high magnetic field is removed. SGR 1099+14 and AXP 1E
1048.1-5937 might be quark stars of $R\sim5$ km and $R\sim$ 2 km,
respectively, under certain assumptions.

The braking index is a good parameter to test torque models (and
emission models) since it is an induced parameter directly from
observation and depends on less assumption. As a result of
observational  difficulties, braking indices of only six
rotation-powered pulsars are obtained \citep[][and references
therein]{Livingstone06}. No AXP/SGR has measured good braking
index due to the high timing noisy. Because accretion would add
additional torque to spin-down, the braking index departs from
$n=3$, e.g. in \citet{Chen06}, varying in the range of $-4$ to $3$
in the accretion model. At the same time, accretion would result
in a frequent change of $\pdot$. Therefore, the braking index
obtained by short-time observation might not be good, to be
contaminated dominantly by timing noise.
Additionally, the braking index of an accreting pulsar could also
be variable, depending on the accretion state. While, good braking
indices of six solitary radio pulsars with no accretion are all in
the range of 1 to 3.

The the accretion torque that we use is simply an approximation.
The detail coupling between accreted matter and magnetosphere is
still not clear. The interaction region could not be a thin layer
at $\rmag$ but have a certain depth. So the thin layer
approximation needs further investigation (via, e.g., numerical
simulation).

\end{document}